\documentclass{ws-p9-75x6-50}
\def\ea{{\it et al.\,}}
\def\eg{{\it e.g., \,}}

\def\rel{relativistic\,}
\def\nrel{nonrelativistic\,}
\def\sz{Sunyaev \& Zeldovich\,}
\def\be{\begin{equation}}
\def\ee{\end{equation}}
\def\cl{\centerline}

\def\bs{\bigskip}

\begin{document}

\title{The Sunyaev-Zeldovich Effect: Current \\Status and Future Prospects}

\author{Yoel Rephaeli}

\address{School of Physics \& Astronomy, Tel Aviv University, Tel Aviv 
69978, Israel\\E-mail: yoelr@wise1.tau.ac.il}
 
\maketitle                             

\abstracts{
The detailed spectral and spatial characteristics of the signature imprinted 
on the cosmic microwave background (CMB) radiation by Compton scattering of 
the radiation by electrons in the hot gas in clusters of galaxies - the 
Sunyaev-Zeldovich (S-Z) effect - are of great astrophysical and 
cosmological significance. In recent years observations of the effect 
have improved tremendously; high signal-to-noise images of the effect 
(at low microwave frequencies) can now be obtained by interferometric 
arrays. In the near future, high frequency measurements of 
the effect will be made with ground based and balloon-borne telescopes 
equipped with bolometeric arrays. Towards the end of the decade the PLANCK 
satellite will carry out an extensive S-Z survey over a wide frequency 
range. Along with the improved observational capabilities, the theoretical 
description of the effect, and its use as a precise cosmological probe, 
have been considerably advanced. In this review, I briefly discuss the 
nature and significance of the effect, its exact theoretical description, 
the current observational status, and prospects for the near future.} 

\section{Introduction}

The S-Z effect was first described by Zeldovich \& Sunyaev (1969) and 
Sunyaev \& Zeldovich (1972), and after many attempts over a decade, 
it was convincingly detected in several clusters by single-dish radio 
telescopes (for general reviews, see Rephaeli 1995a, Birkinshaw 1999). 
Increased realization of the cosmological significance of the effect has 
led to major improvements in observational techniques and to extensive 
theoretical investigations of many of its facets. The most important 
recent development is the ability to carry out sensitive radio 
interferometric measurements of the effect (Jones 1993, Carlstrom \ea 
1996). Impressive images of the effect in more than thirty clusters have 
already been obtained using the OVRO and BIMA arrays (Carlstrom \ea 1999, 
Carlstrom \ea 2000). Theoretical treatment of the S-Z effect has also 
improved, starting with the work of Rephaeli (1995b), who performed 
an exact relativistic calculation and demonstrated the need for such a 
more accurate description. 

The effect is a direct probe of clusters; high resolution measurements 
yield the spatial distributions of the hot intracluster (IC) gas and 
the {\it total} mass, as well as information on the evolution of 
clusters. Of more basic importance is the ability to determine the Hubble 
constant, $H_0$, and the density parameter, $\Omega$, from S-Z and X-ray 
measurements. This method to determine $H_0$, which has clear advantages 
over the traditional galactic distance ladder method, has already been 
employed but its full potential has not yet been realized because of 
substantial observational uncertainties. Sensitive mapping of the effect,  
and the understanding and minimization of systematic uncertainties in the 
S-Z and X-ray measurements, are essential steps towards this goal and 
constitute the main challenge of current and near future work on the S-Z 
effect.

The recent progress in S-Z work, and especially the improved spectral 
and spatial capabilities that will be attained in the near future by 
ground based and stratospheric projects, will advance this field to 
the forefront of cosmological research. Here I review recent progress 
in this field and briefly discuss the prospects for the near future.

\section{Exact Description of the Effect}

Compton scattering of the CMB by hot gas heats the radiation, resulting 
in a systematic transfer of photons from the Rayleigh-Jeans (R-J) to the 
Wien side of the (Planckian) spectrum. An accurate description 
of the interaction of the radiation with a hot electron gas necessitates 
the calculation of the exact frequency re-distribution function in the 
context of a relativistic formulation. The calculations of Sunyaev \& 
Zeldovich (1972) are based on a solution to the {\it \nrel} Kompaneets 
(1957) equation. The result of their treatment is a simple expression for 
the intensity change resulting from the scattering off electrons with 
thermal velocity distribution, $\Delta I_t$, in terms of the 
Comptonization parameter, $y = \int (kT_e/mc^2) n \sigma _T dl$. This is 
a line of sight integral (through the cluster) over the electron density 
($n$) and temperature ($T_e$); $\sigma _T$ is the Thomson cross section.
$\Delta I_t$ 
is negative in the R-J region and positive at frequencies above a critical 
value $\sim 217$ GHz ($x\equiv h\nu/kT = 3.83$). Typically, in a rich cluster, 
$y \sim 10^{-4}$, along a line of sight through the center, and the 
magnitude of the relative temperature change due to the thermal effect 
is $\Delta T_t/T_0 = -2y \sim -2 \cdot 10^{-4}$ in the R-J region.  

The effect has a second component when the cluster has a finite (peculiar) 
velocity in the CMB frame. This kinematic Doppler component, which is 
obviously proportional to the component of the cluster peculiar velocity 
along the line of sight, $v_r$, and to the Thomson optical depth, is 
$\Delta I_k = x^{4}e^{x}(v_r/c)\tau /(e^x -1)^2$. The related temperature 
change is $\Delta T_k/T_0 = -(v_r/c)\tau$ (Sunyaev \& Zeldovich 1980). 

The \nrel treatment of Sunyaev \& Zeldovich (1972) is generally valid at 
low gas temperatures and at low frequencies. Rephaeli (1995b) has shown that 
this approximation is insufficiently accurate for use of the effect 
as a precise cosmological probe: Electron velocities in the IC gas are 
high, and the relative photon energy change in the scattering is appreciable 
enough to require a \rel calculation. Using the exact probability 
distribution in Compton scattering, and the relativistically correct 
form of the electron Maxwellian velocity distribution, Rephaeli (1995b) 
calculated $\Delta I_t$ in the limit of small $\tau$, keeping terms 
linear in $\tau$ (single scattering). Results of this semi-analytic 
calculation, shown in Figure 1, demonstrate that the relativistic spectral 
distribution of the intensity change is quite different from that derived 
by Sunyaev \& Zeldovich (1972). Deviations from their expression increase 
with $T_e$ and can be quite substantial. These are especially large near the 
crossover frequency (where $\Delta T_t$ changes sign) which shifts to higher 
values with increasing gas temperature. 

\begin{figure}[t]
\hspace*{0.4in}{\psfig{file=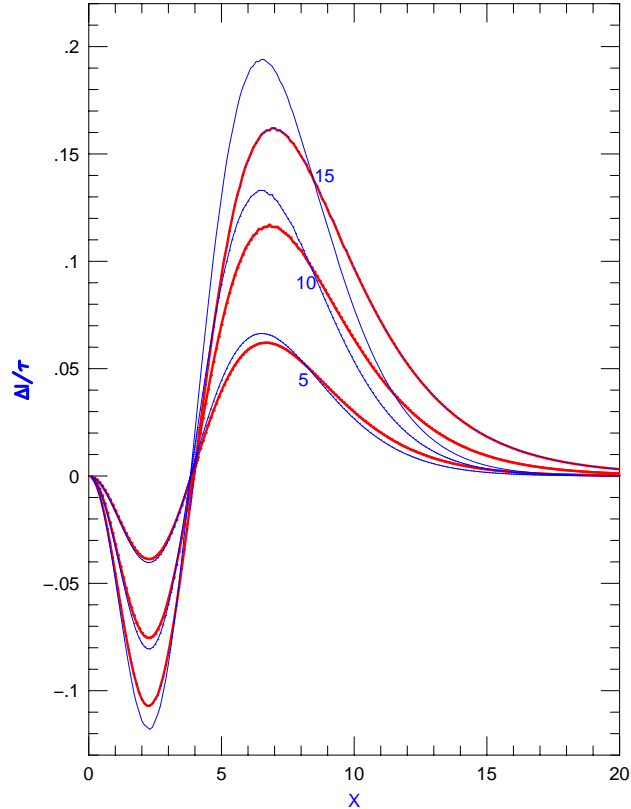, width=10cm, angle=270}}
\bs
\bs
\bs
\bs

\caption{The spectral distribution of $\Delta I_t/\tau$ [in units 
of $(hc)^2/(2 (kT_o)^3)$]. The pairs of thick (red) and thin (blue) lines 
show the \rel and nonrelativistic distributions, respectively. Three pairs 
of lines are shown corresponding to $kT_e =$ 5, 10, and 15 keV.}
\end{figure}
       
The work of Rephaeli (1995b) sparked considerable interest which led 
to various generalizations and extensions of the relativistic treatment of 
the S-Z effect. Challinor \& Lasenby (1998) generalized the Kompaneets 
equation and obtained analytic approximations to its solution for the 
change of the photon occupation number by means of a power series 
in $\theta_{e}$ = $kT_{e}/mc^{2}$. Itoh \ea (1998) adopted this 
approach and improved the accuracy of the analytic approximation by 
expanding to fifth order in $\theta_{e}$. Sazonov \& Sunyaev (1998) 
and Nozawa \ea (1998) have extended the relativistic treatment also to 
the kinematic component obtaining -- for the first time -- the leading 
cross terms in the expression for the total intensity change ($\Delta I 
= \Delta I_t + \Delta I_k$) which depend on both $T_e$ and $v_r$. An 
improved analytic fit to the results of numerical integration (of the 
collision term in the Boltzmann equation), valid for $0.02 \leq 
\theta_{e} \leq 0.05$, and $x \leq 20$ ($\nu \leq 1130$ GHz), has recently 
been given by Nozawa \ea (2000). In view of the possibility that in some 
rich clusters $\tau \sim 0.02-0.03$, the approximate analytic expansion 
to fifth order in $\theta_{e}$ necessitates self-consistent inclusion 
also of multiple scatterings, of order $\tau^2$. This has been 
accomplished by Itoh \ea (2000), and Shimon \& Rephaeli (2001).

The relativistically correct calculation has to be used in all 
high frequency S-Z work, especially when measurements of the effect 
are used to determine precise values of the cosmological parameters. 
Also, since the ability to determine peculiar velocities of clusters 
depends very much on measurements at or very near the crossover frequency, 
its exact value has to be known. This necessitates knowledge of the gas 
temperature since in the exact \rel treatment the crossover frequency is 
no longer independent of the gas temperature, and is approximately given 
by $\simeq 217[1+1.167 kT_e/mc^2 -0.853(kT_e/mc^2)^2]$ GHz (Nozawa \ea 1998a).

Use of the S-Z effect as a cosmological probe necessitates also X-ray 
measurements to determine (at least) the gas temperature. Therefore, a 
relativistically correct expression for the (spectral) bremsstrahlung 
emissivity must be used (Rephaeli \& Yankovitch 1997). In the latter paper 
first order relativistic corrections to the velocity distribution, and 
electron-electron bremsstrahlung, were taken into account in correcting 
values of $H_0$ that were previously derived using the \nrel expression 
for the emissivity (see also Hughes \& Birkinshaw 1998). Nozawa \ea (1998b) 
have performed a more exact calculation of the \rel bremsstrahlung Gaunt 
factor.

Compton scattering of the CMB in clusters also affects its polarization 
towards the cluster. Net polarization is induced due to the quadrupole 
component in the spatial distribution of the radiation, and when the 
cluster peculiar velocity has a component transverse to the line of sight, 
$v_{\perp}$ (\sz 1980, Sazonov \& Sunyaev 1999). The leading contributions 
to the latter, kinematically-induced polarization, are proportional to 
$(v_{\perp}/c) \tau^2$ and $(v_{\perp}/c)^2\tau$. Itoh \ea (2000) have 
included relativistic corrections in the expression for the kinematic 
polarization.

\section{Recent Measurements}

The dramatic improvements in the quality of S-Z measurements over the last 
seven years is largely due to the use of interferometric arrays. Dish 
arrays have several major advantages over a single dish, including 
insensitivity to changes in the atmospheric emission, sensitivity to 
specific angular scales and to signals which are correlated between the 
array elements, and the high angular resolution that enables nearly 
optimal subtraction of signals from point sources. With the improved 
sensitivity of radio receivers it became 
advantageous to use interferometric arrays for S-Z imaging measurements 
(starting with the use of the Ryle telescope by Jones \ea 1993). Current 
state-of-the-art work is done with the BIMA and OVRO arrays; images of 
some 35 moderately distant  clusters (in the redshift range $0.17 - 
0.89$) have already been obtained at $\sim 30$ GHz (Carlstrom \ea 1999, 2000). 
A beautiful example is the 28 GHz image of the cluster CL0016+16 obtained 
with the BIMA array (Carlstrom \ea 1999) which is shown in Figure 2. The 
ROSAT X-ray image is superposed on the contour lines of the S-Z profile 
in the lower frame; note the good agreement between the orientation of 
the X-ray and S-Z brightness distributions. 

\begin{figure}
\cl{\psfig{file=fig2_1.epsi, width=10cm, angle=270}}

\hspace*{1.6in}{{\psfig{file=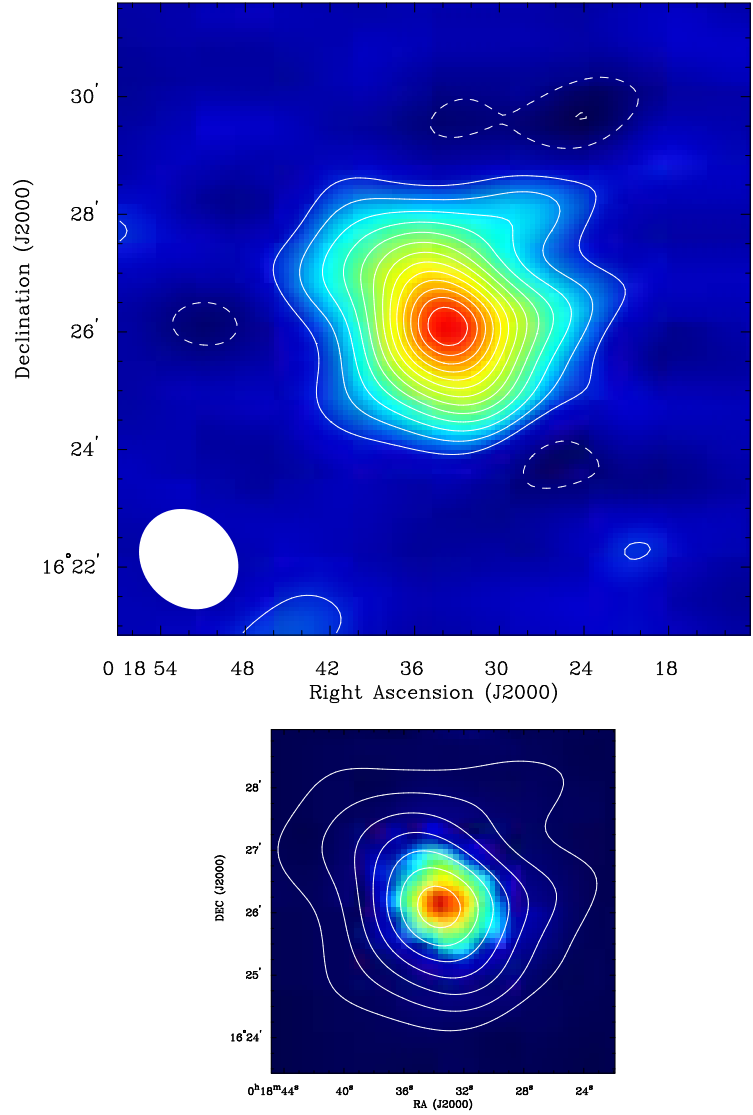, width=5.8cm, angle=0}}}
\caption{\small{S-Z and X-ray images of the cluster CL0016+16. The S-Z
image (false color) of the cluster, obtained with the BIMA array, is shown
in the upper frame. In the lower frame, contours of the S-Z effect in the
cluster are superposed on the {\it ROSAT} X-ray (false color) image (from
Carlstrom \ea 1999).}}
\end{figure}  

Work has begun recently with the CBI, a new radio (26-36 GHz) 
interferometric array (in the Chilean Andes) of small (0.9m) dishes on a 
platform with baselines in the 1m to 6m range. The spatial resolution of 
the CBI is in the $3'-10'$ range, so (unlike the BIMA and OVRO arrays) it is 
suitable for S-Z measurements of nearby clusters. Some 9 clusters have 
already been observed with the CBI (Udomprasert \ea 2000).
        
Recent work at higher frequencies includes measurements with the SuZIE 
array, and the PRONAOS and MITO telescopes. SuZIE (which was previously 
used to measure the effect towards the moderately distant clusters A1689 
\& A2163 Holzapfel \ea 1997a, 1997b) observed the effect towards A1835 at 
three spectral bands centered on 145, 221, 279 GHz (Mauskopf \ea 2000). 
Observations at four wide spectral bands (in the overall range of 285-1765 
GHz) were made of A2163 with the PRONAOS atmospheric 2m telescope (Lamarre 
\ea 1998), leading to what seems to be the first detection of the effect 
by a balloon-borne experiment. The new 2.6m MITO telescope -- which 
currently operates at four spectral bands and has a large $\sim 17'$ 
beam -- was used to observe the effect in the Coma cluster (Lamagna \ea, 
these proceedings). 
The strongest S-Z effect, with $y = 1.2\times 10^{-3}$, seems to have been 
detected towards the distant (z=0.45) cluster RXJ 1347 (Pointecouteau \ea 
1999) with the Diabolo bolometer operating at the IRAM 30m radio 
telescope. The Diabolo has a $0.5'$ beam, 
and a dual channel bolometer (centered on 2.1 and 1.2 mm). 
Four other clusters were also observed with diabolo (Desert \ea 1998).

\section{The Effect as a Cosmological Probe}

The S-Z effect is a unique probe whose indispensable qualities are being 
increasingly exploited for the determination of cluster and global 
cosmological parameters. For an extensive discussion of use of the effect 
as a probe, see the reviews by Rephaeli (1995a), and Birkinshaw (1999). 

In principle, high spatial resolution measurements of the S-Z effect yield 
the gas temperature and density profiles across the cluster. The deduction 
of an approximate 3D density and temperature distributions from their 
sky-projected profiles requires use of a deprojection algorithm. One such 
was developed by Zaroubi \ea (1998; see also the contribution by Zaroubi 
to these proceedings). Cluster gas density and temperature profiles have 
been mostly deduced from X-ray measurements. S-Z measurements can more 
fully determine these profiles due to the linear dependence of $\Delta 
I_t$ on $n$, as compared to the $n^2$ dependence of the (thermal 
bremsstrahlung) X-ray brightness profile. 

The cluster full mass profile, $M(r)$, can be derived directly from the 
gas density and temperature distributions by solving the equation of 
hydrostatic equilibrium (assuming, of course, the gas has reached such a 
state in the underlying gravitational potential). This method has already 
been employed in many analyses using X-ray deduced gas parameters (\eg 
Fabricant \ea 1980). Grego \ea (2001) have recently used this method 
to determine total masses and gas mass fractions of 18 clusters based 
largely on the results of their interferometric S-Z measurements. 
Isothermal gas with the familiar density profile, 
$(1 + r^2/r_c^2)^{-3\beta/2}$, was assumed. The core radius, $r_c$, and 
$\beta$ were determined from analysis of the S-Z data, whereas the X-ray 
value of the temperature was adopted. Mean values of the gas mass fraction 
were found to be in the range $(0.06-0.09)h^{-1}$ (where $h$ is the value 
of $H_0$ in units of $100$ km s$^{-1}$ Mpc$^{-1}$) for the currently popular 
open and flat, $\Lambda$-dominated, CDM models.

Measurement of the kinematic S-Z effect yields the line of sight 
component of the cluster peculiar velocity ($v_r$). This is 
observationally feasible only in a narrow spectral band near the
critical frequency, where the thermal effect vanishes while the kinematic
effect -- which is usually swamped by the much larger thermal component --
is maximal (Rephaeli \& Lahav 1991). SuZIE is the first experiment with 
a spectral band centered on the crossover frequency. Measurements of 
the clusters A1689 and A2163 (Holzapfel \ea 1997b) and A1835 (Mauskopf 
\ea 2000) yielded substantially uncertain results for $v_r$ 
($170^{+815}_{-630}$, $490^{+1370}_{-880}$, and $500 \pm 1000$ km s$^{-1}$, 
respectively).

Perhaps the most important use of the S-Z effect so far has been the 
measurement of the Hubble constant, $H_0$, and the cosmological density 
parameter, $\Omega$. Briefly, the method is based on determining the 
angular diameter distance, $d_A$, from a combination of $\Delta I_t$, the 
X-ray surface brightness, and their spatial profiles. Averaging over 
the first eight determinations of $H_0$ (from S-Z and X-ray measurements 
of seven clusters) yielded $H_0 \simeq 58 \pm 6$ km s$^{-1}$ Mpc$^{-1}$ 
(Rephaeli 1995a), but the database was very non-uniform and the 
errors did not include systematic uncertainties. Repeating this with a 
somewhat updated data set, Birkinshaw (1999) deduced essentially a similar 
mean value.
A much larger S-Z data set is now available from the interferometric 
BIMA and OVRO observations, and since the redshift range of the clusters 
in the sample is substantial, the dependence on $\Omega$ is appreciable. 
A fit to 33 cluster distances gives $H_0 = 60$ km s$^{-1}$ Mpc$^{-1}$ for 
$\Omega = 0.3$, and $H_0 = 58$ km s$^{-1}$ Mpc$^{-1}$ for $\Omega = 1$, 
with direct observational errors of $\pm 5\%$ (Carlstrom \ea 2000). The 
main known sources of systematic uncertainties (see discussions in Rephaeli 
1995a, and Birkinshaw 1999) are presumed to introduce an additional error 
of $\sim 30\%$ (Carlstrom \ea 2000). The current number of clusters with S-Z 
determined distances is sufficiently large that a plot of $d_A$ vs. 
redshift (a Hubble diagram) is now quite of interest, but with the present 
level of uncertainties the limits on the value of $\Omega$ are not very 
meaningful (as can be seen from figure 11 of Carlstrom \ea 2000).

CMB anisotropy induced by the S-Z effect (Sunyaev 1977, Rephaeli 1981) 
is the main source of secondary anisotropy on angular scales of few 
arcminutes. Because of this, and the great interest in this range of 
angular scales -- multipoles (in the representation of the CMB 
temperature in terms of spherical harmonics) $\ell \geq 1000$ -- 
the S-Z anisotropy has been studied extensively in the last few years. 
The basic goal has been to map its predicted $\ell$ dependence in 
viable cosmological, large scale structure, and IC gas models. The 
strong motivation for this is the need to accurately calculate the 
power spectrum of the full anisotropy in order to make precise global 
parameter determinations from large stratospheric and satellite 
databases. In addition, mapping the S-Z anisotropy will yield direct 
information on the evolution of the cluster population.

Results from many calculations of the predicted S-Z anisotropy are not 
always consistent even for the same cosmological and large scale 
structure parameters. This is simply due to the fact that the calculation 
involves a large number of input parameters in addition to the global 
cosmological parameters (\eg the present cluster density, and parameters 
characterizing the evolutionary history of IC gas), and the 
sensitive dependence of the anisotropy on some of these. 
The anisotropy can also be predicted -- presumably more directly -- from 
simulations of the S-Z sky based largely on results from cluster X-ray 
surveys and the use of simple scaling relations (first implemented by 
Markevitch \ea 1992).

The anticipated observational capabilities, to be achieved with 
long duration balloon-borne experiments and satellites, of detailed mapping 
of the small angular scale anisotropy, have motivated many recent works. 
Colafrancesco \ea (1997), and Kitayma \ea (1998), have calculated the S-Z 
cluster number counts in an array of open and flat cosmological and dark 
matter models, and Seljak \ea (2000) have recently carried out 
hydrodynamical simulations in order to generate S-Z maps and power spectra. 
Cooray \ea (2000) have, in particular, concluded 
that the planned multi-frequency survey with the Planck satellite should 
be able to distinguish between the primary and S-Z anisotropies, and 
measure the latter with sufficient precision to determine its power 
spectrum and higher order correlations.

The main characteristics of the predicted power spectrum of the induced 
S-Z anisotropy are shown in Figure 3. Plotted are the angular power spectrum 
functions, $C_{\ell}(\ell +1)/2\pi$, vs. the multipole, $\ell$, for both 
the primary and S-Z induced anisotropies in the standard CDM model with 
$\Omega =1$, based on the work of Sadeh and Rephaeli (2001). Their treatment 
is an extension of the approach adopted by Colafrancesco \ea (1997), who 
used the Press \& Schechter formulation for the calculation of the cluster 
density as function of mass and redshift, and normalized it at the current 
epoch by the observed X-ray luminosity function (see also Colafrancesco \ea 
1994). IC gas was assumed to evolve in a simple manner consistent with the 
results of the EMSS survey carried out with the Einstein satellite. The 
primary anisotropy was calculated using the CMBFAST code of Seljak \& 
Zaldarriaga (1996). The solid line shows the primary anisotropy which dominates 
over the S-Z anisotropy for $\ell < 3000$. The S-Z power is largely due to 
the thermal effect; this rises with $\ell$ and is maximal around  $\ell 
\sim 1000$.
In this model, the S-Z power spectrum contributes a fraction of 5\% (10\%) 
to the primary anisotropy at $\ell \simeq 1860$ ($\ell \simeq 2360$). The 
relative magnitude of the S-Z power is higher in the $\Lambda$-dominated 
CDM model with $\Lambda = 0.8$; 5\% and 10\% contributions occur at 
$\ell \simeq 1520$ and $\ell \simeq 1840$, respectively. 
It can be concluded from this (and other studies) that the S-Z induced 
anisotropy has to be taken into account if the extraction of the cosmological 
parameters from an analysis of measurements of the CMB power spectrum at 
$\ell > 1300$ is to be precise.

\begin{figure}[t]
\cl{\psfig{file=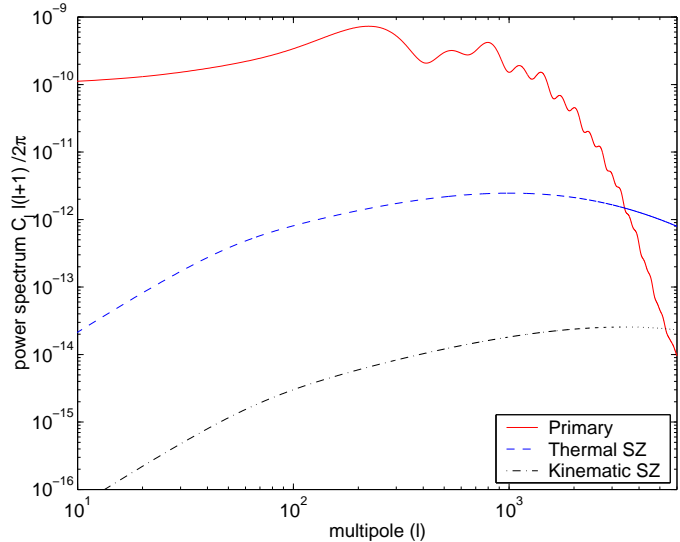, width=9cm, angle=0}}

\caption{\small{Primary and S-Z power spectra in the flat CDM model (Sadeh \& 
Rephaeli 2001). The solid (red) line shows the primary anisotropy as calculated 
using the CMBFAST computer code of Seljak \& Zaldarriaga (1996). The dashed 
(blue) line shows the thermal S-Z power spectrum, and the dotted-dashed 
(black) line is the contribution of the kinematic component.}}
\end{figure}   

\section{Future Prospects}

Work on the S-Z effect has greatly advanced in the last few years, and the 
prospects are very good for major improvements in the near future. The 
full potential of the effect as an important cosmological probe will be 
realized when high quality spectral and spatial mapping of the effect 
in {\it nearby} clusters, $z \leq 0.1$, will be feasible. The main 
limitation on the accuracy of the cosmological parameters will continue to 
be due to systematic uncertainties, which can be most optimally controlled 
when sensitive S-Z and X-ray measurements are made of nearby clusters. 
The interferometric CBI array extended the capability of the BIMA and OVRO 
arrays to obtain sensitive maps of the effect in nearby clusters. The 
currently operational XMM and {\it Chandra} satellites provide 
state-of-the-art X-ray spectral and spatial capabilities. Use of the S-Z 
spectrum as a powerful diagnostic tool will soon be possible when sensitive 
multi-frequency bolometric array experiments, with spectral bands in the 
range $150-350$ GHz, and very good spatial resolution ($\sim 3'$), become 
operational. Both ground-based (\eg the upgraded MITO experiment) and 
stratospheric (\eg BOOST, OLIMPO) projects are underway. Measurements of a 
large number of clusters with these and the currently operational arrays 
will enable a much more precise determination of the cosmological parameters. 
In particular, it will be possible to measure $H_0$ with an overall uncertainty 
of just $\sim 5\%$.

\end{document}